\documentclass[
a4paper,
oneside,
dvips,
11pt]{book}

\newif\ifpdf\ifx\pdfoutput\undefined\pdffalse\else\pdfoutput=1\pdftrue\fi

\usepackage{hyperref}
\usepackage{epsfig}                             
\usepackage{geometry}                           
\usepackage{verbatim}                           
\usepackage{amssymb}                            
\usepackage{caption2}                     
\usepackage{subfigure}
\usepackage{booktabs}                           
\usepackage{tabularx}                           
\usepackage{rotating}
\usepackage{lscape}
\usepackage{ifthen}
\usepackage{xspace}
\usepackage{setspace}
\begin{document}
\addcontentsline{toc}{chapter}{Summary}%

\pagestyle{plain}

\vspace*{1cm}

{\Huge Lucky Exposures: Diffraction limited astronomical imaging through
  the atmosphere}
\vspace*{1cm}

{\huge Ph.D. Thesis Dissertation by \href{http://www.strw.leidenuniv.nl/\%7Etubbs/contact/}{Robert Nigel Tubbs}}
\vspace*{1cm}

{\large \href{http://www.mrao.cam.ac.uk/telescopes/coast/}{COAST Group}, Cambridge University, UK}
\vspace*{1cm}

{\large September 2003}
\vspace*{1cm}

{\Huge Thesis Summary}
\vspace*{1cm}

The resolution of astronomical imaging from large optical telescopes
is usually limited by the blurring effects of refractive index
fluctuations in the Earth's atmosphere. By taking a large number of
short exposure images through the atmosphere, and then selecting,
re-centring and co-adding the best images this resolution limit can be
overcome. This approach has significant benefits over other techniques
for high-resolution optical imaging from the ground. In particular the
reference stars used for our method (the \emph{Lucky Exposures}
technique) can generally be fainter than those required for the
natural guide star adaptive optics approach or those required for
other speckle imaging techniques. The low complexity and low
instrumentation costs associated with the Lucky Exposures method make
it appealing for medium-sized astronomical observatories.

The method can provide essentially diffraction-limited I-band imaging
from well-figured ground-based telescopes as large as $2.5$ $m$
diameter. The faint limiting magnitude and large isoplanatic patch
size for the Lucky Exposures technique at the Nordic Optical Telescope
means that $25\%$ of the night sky is within range of a suitable
reference star for I-band imaging. Typically the $1\%$---$10\%$ of
exposures with the highest Strehl ratios are selected. When these
exposures are shifted and added together, field stars in the resulting
images have Strehl ratios as high as $0.26$ and full width at half
maximum flux (FWHM) as small as $90$ $milliarcseconds$. Within the
selected exposures the isoplanatic patch is found to be
$60$ $arcseconds$ in diameter at $810$ $nm$ wavelength. Images within
globular clusters and of multiple stars from the Nordic Optical
Telescope using reference stars as faint as $I\sim16$ are presented.

A new generation of CCDs (\href{http://e2vtechnologies.com/}{E2V} L3Vision CCDs) were used in
these observations, allowing extremely low noise high frame-rate
imaging with both fine pixel sampling and a relatively wide field of
view. The theoretical performance of these CCDs is compared with the
experimental results obtained.

\clearpage
\thispagestyle{empty}

\chapter*{Download options}
\addcontentsline{toc}{chapter}{Download Thesis}%

{\huge PDF version of thesis dissertation with full quality figures (\textbf{Recommended})}
\begin{itemize}
\item \href{http://www.strw.leidenuniv.nl/\%7Etubbs/thesis/rnt_thesis.pdf}{http://www.strw.leidenuniv.nl/$\sim$tubbs/thesis/rnt{\_}thesis.pdf}
- PDF version of thesis dissertation (18 MB)
\item \href{http://www.mrao.cam.ac.uk/telescopes/coast/theses/rnt/pdf_files/rnt_thesis.zip}{http://www.mrao.cam.ac.uk/telescopes/coast/theses/rnt/pdf{\_}files/rnt{\_}thesis.zip} - zipped version of
the same PDF file,
compatible with PKZIP and LINUX zip (8 MB)
\end{itemize}
Lower quality versions of individual chapters with smaller PDF file sizes
can be downloaded from the HTML version of the thesis below.

\vspace*{1cm}
{\huge Lower quality versions with smaller file download sizes}
\vspace*{0.5cm}

\textbf{Note that many figures are poorly reproduced in these versions of
the document. To view a full quality document, download one of the
PDF versions listed above
}

\vspace*{0.5cm}
\href{http://www.strw.leidenuniv.nl/\%7Etubbs/thesis/contents.html}{Access full table of contents for the HTML version
of the thesis}

\begin{itemize}
\item \textbf{\href{http://www.mrao.cam.ac.uk/telescopes/coast/theses/rnt/thesis.html}{\emph{Front Matter} from thesis}}
\item \textbf{\href{http://www.mrao.cam.ac.uk/telescopes/coast/theses/rnt/node2.html}{Introduction}} (also available as a \href{http://www.mrao.cam.ac.uk/telescopes/coast/theses/rnt/pdf_files/chapter1.pdf}{low quality version in PDF format} - 600kB)
\item \textbf{\href{http://www.mrao.cam.ac.uk/telescopes/coast/theses/rnt/node22.html}{Lucky Exposures}} (also available as a \href{http://www.mrao.cam.ac.uk/telescopes/coast/theses/rnt/pdf_files/chapter2.pdf}{low quality version in PDF format} - 400kB)
\item \textbf{\href{http://www.mrao.cam.ac.uk/telescopes/coast/theses/rnt/node34.html}{Observations of
bright sources at the NOT }} (also available as a \href{http://www.mrao.cam.ac.uk/telescopes/coast/theses/rnt/pdf_files/chapter3.pdf}{low quality version in PDF format} - 2MB)
\item \textbf{\href{http://www.mrao.cam.ac.uk/telescopes/coast/theses/rnt/node59.html}{Electron multiplying
CCD performance}} (also available as a \href{http://www.mrao.cam.ac.uk/telescopes/coast/theses/rnt/pdf_files/chapter4.pdf}{low quality version in PDF format} - 800kB)
\item \textbf{\href{http://www.mrao.cam.ac.uk/telescopes/coast/theses/rnt/node80.html}{Observations
with a low noise CCD at the NOT}} (also available as a \href{http://www.mrao.cam.ac.uk/telescopes/coast/theses/rnt/pdf_files/chapter5.pdf}{low quality version in PDF format} - 2MB)
\item \textbf{\href{http://www.mrao.cam.ac.uk/telescopes/coast/theses/rnt/node99.html}{Conclusions}} (also available as a \href{http://www.mrao.cam.ac.uk/telescopes/coast/theses/rnt/pdf_files/chapter6.pdf}{low quality version in PDF format} - 50kB)
\item \textbf{\href{http://www.mrao.cam.ac.uk/telescopes/coast/theses/rnt/node107.html}{Simplified approximation to a single Taylor screen atmosphere}} (also available as a \href{http://www.mrao.cam.ac.uk/telescopes/coast/theses/rnt/pdf_files/appendixa.pdf}{low quality version in PDF format} - 100kB)
\item \textbf{\href{http://www.mrao.cam.ac.uk/telescopes/coast/theses/rnt/node108.html}{Observation log for June-July 2003 observations at the NOT }} (also available as a \href{http://www.mrao.cam.ac.uk/telescopes/coast/theses/rnt/pdf_files/appendixb.pdf}{low quality version in PDF format} - 60kB)
\item \textbf{\href{http://www.mrao.cam.ac.uk/telescopes/coast/theses/rnt/node109.html}{Bibliography}} (also available as a \href{http://www.mrao.cam.ac.uk/telescopes/coast/theses/rnt/pdf_files/bibliography.pdf}{low quality version in PDF format} - 110kB)
\end{itemize}

\clearpage
\thispagestyle{empty}

\chapter*{Bibliography}
\addcontentsline{toc}{chapter}{Bibliography}%

A{\small IME}, C., R{\small ICORT}, G., K{\small ADIRI}, S., \& M{\small ARTIN}, F. 1981.

  \href{http://ukads.nottingham.ac.uk/cgi-bin/nph-bib_query?bibcode=1981OptCo..39..287A\&db_key=AST}{Temporal autocorrelation functions of solar speckle
  patterns}.
\newline \emph{Optics Communications}, \textbf{39}(Nov.), 287-292.

\vspace*{0.25cm}
A{\small IME}, C., B{\small ORGNINO}, J., M{\small ARTIN}, F., P{\small ETROV}, R., \& R{\small ICORT}, G.
  1986.

  \href{http://ukads.nottingham.ac.uk/cgi-bin/nph-bib_query?bibcode=1986OSAJ....3.1001A\&db_key=AST}{Contribution to the space-time study of stellar speckle
  patterns}.
\newline \emph{Optical Society of America, Journal, A: Optics and Image
  Science}, \textbf{3}(July), 1001-1009.

\vspace*{0.25cm}
A{\small LLER}, L.~H., A{\small PPENZELLER}, I., B{\small ASCHEK}, B., D{\small UERBECK}, H.~W.,
  H{\small ERCZEG}, T., L{\small AMLA}, E., M{\small EYER-}H{\small OFMEISTER}, E., S{\small CHMIDT-}K{\small ALER}, T.,
  S{\small CHOLZ}, M., S{\small EGGEWISS}, W., S{\small EITTER}, W.~C., \& W{\small EIDEMANN}, V. (eds).
  1982.
 \emph{  \href{http://ukads.nottingham.ac.uk/cgi-bin/nph-bib_query?bibcode=1982lbor.book.....A\&db_key=AST}{Landolt-Börnstein: Numerical Data and Functional
  Relationships in Science and Technology - Group VI (Astronomy and
  Astrophysics), Vol. 2b (Stars and Star Clusters), Schaifers \& Voigt, Eds.,
  Chapter 4.2}}.

\vspace*{0.25cm}
A{\small NDERSEN}, M.~I., \& S{\small {\O}RENSEN}, A.~N. 1996 (Feb.).
 \emph{Image quality at the Nordic Optical Telescope}.
\newline IJAF Report~2. Copenhagen University Observatory, Denmark.

\vspace*{0.25cm}
A{\small VILA}, R., V{\small ERNIN}, J., \& M{\small ASCIADRI}, E. 1997.

  \href{http://ukads.nottingham.ac.uk/cgi-bin/nph-bib_query?bibcode=1997ApOpt..36.7898A\&db_key=INST}{Whole atmospheric-turbulence profiling with generalized
  scidar}.
\newline \emph{Applied Optics}, \textbf{36}(Oct.), 7898-7905.

\vspace*{0.25cm}
B{\small AHCALL}, J.~N., \& S{\small ONEIRA}, R.~M. 1984.

  \href{http://ukads.nottingham.ac.uk/cgi-bin/nph-bib_query?bibcode=1984ApJS...55...67B\&db_key=AST}{Comparisons of a standard galaxy model with stellar
  observations in five fields}.
\newline \emph{Astrophysical Journal Supplement Series}, \textbf{55}(May), 67-99.

\vspace*{0.25cm}
B{\small ALDWIN}, J.~E., H{\small ANIFF}, C.~A., M{\small ACKAY}, C.~D., \& W{\small ARNER}, P.~J.
  1986.

  \href{http://ukads.nottingham.ac.uk/cgi-bin/nph-bib_query?bibcode=1986Natur.320..595B\&db_key=AST}{Closure phase in high-resolution optical imaging}.
\newline \emph{Nature}, \textbf{320}(Apr.), 595-597.

\vspace*{0.25cm}
B{\small ALDWIN}, J.~E., B{\small ECKETT}, M.~G., B{\small OYSEN}, R.~C., B{\small URNS}, D.,
  B{\small USCHER}, D.~F., C{\small OX}, G.~C., H{\small ANIFF}, C.~A., M{\small ACKAY}, C.~D.,
  N{\small IGHTINGALE}, N.~S., R{\small OGERS}, J., S{\small CHEUER}, P.~A.~G., S{\small COTT}, T.~R.,
  T{\small UTHILL}, P.~G., W{\small ARNER}, P.~J., W{\small ILSON}, D.~M.~A., \& W{\small ILSON}, R.~W.
  1996.

  \href{http://ukads.nottingham.ac.uk/cgi-bin/nph-bib_query?bibcode=1996A\%26A...306L..13B\&db_key=AST}{The first images from an optical aperture synthesis
  array: mapping of Capella with COAST at two epochs.}
\newline \emph{Astronomy and Astrophysics}, \textbf{306}(Feb.), L13+.

\vspace*{0.25cm}
B{\small ALDWIN}, J.~E., T{\small UBBS}, R.~N., C{\small OX}, G.~C., M{\small ACKAY}, C.~D., W{\small ILSON},
  R.~W., \& A{\small NDERSEN}, M.~I. 2001.

  \href{http://ukads.nottingham.ac.uk/cgi-bin/nph-bib_query?bibcode=2001A\%26A...368L...1B\&db_key=AST}{Diffraction-limited 800 nm imaging with the 2.56 m
  Nordic Optical Telescope}.
\newline \emph{Astronomy and Astrophysics}, \textbf{368}(Mar.), L1-L4 (\href{http://www.arxiv.org/abs/astro-ph/0101408}{astro-ph/0101408}).

\vspace*{0.25cm}
B{\small ARNABY}, D., S{\small PILLAR}, E., C{\small HRISTOU}, J.~C., \& D{\small RUMMOND}, J.~D.
  2000.

  \href{http://ukads.nottingham.ac.uk/cgi-bin/nph-bib_query?bibcode=2000AJ....119..378B\&db_key=AST}{Measurements of Binary Stars with the Starfire Optical
  Range Adaptive Optics Systems}.
\newline \emph{Astronomical Journal}, \textbf{119}(Jan.), 378-389.

\vspace*{0.25cm}
B{\small ASDEN}, A.~G., H{\small ANIFF}, C.~A., \& M{\small ACKAY}, C.~D. 2003.

  \href{http://ukads.nottingham.ac.uk/cgi-bin/nph-bib_query?bibcode=2003MNRAS.345..985}{Photon counting strategies with low light level CCDs}.
\newline \emph{Monthly Notices of the Royal Astronomical Society},
  \textbf{345}(Nov.), 985-991 (\href{http://www.arxiv.org/abs/astro-ph/0307305}{astro-ph/0307305}).

\vspace*{0.25cm}
B{\small ATCHELOR}, G.~K., \& T{\small OWNSEND}, A.~A. 1949 (May).
 The nature of turbulent motion at large wave-numbers.
\newline \emph{Pages  238-255 of:} \emph{Proceedings of the Royal Society of
  London A, 199}.

\vspace*{0.25cm}
B{\small URNS}, D., B{\small ALDWIN}, J.~E., B{\small OYSEN}, R.~C., H{\small ANIFF}, C.~A., L{\small AWSON},
  P.~R., M{\small ACKAY}, C.~D., R{\small OGERS}, J., S{\small COTT}, T.~R., W{\small ARNER}, P.~J.,
  W{\small ILSON}, D.~M.~A., \& Y{\small OUNG}, J.~S. 1997.

  \href{http://ukads.nottingham.ac.uk/cgi-bin/nph-bib_query?bibcode=1997MNRAS.290L..11B\&db_key=AST}{The surface structure and limb-darkening profile of
  Betelgeuse}.
\newline \emph{Monthly Notices of the Royal Astronomical Society}, \textbf{290}(Sept.), L11-L16.

\vspace*{0.25cm}
B{\small URT}, D., \& B{\small ELL}, R. 1998.
 CCD imagers with multiplication register.
\newline \emph{European Patent Application Bulletin}, \textbf{39}(Sept.), EP 0 866
  501 A1.

\vspace*{0.25cm}
B{\small USCHER}, D.~F., \& H{\small ANIFF}, C.~A. 1993.

  \href{http://ukads.nottingham.ac.uk/cgi-bin/nph-bib_query?bibcode=1993OSAJ...10.1882B\&db_key=INST}{Diffraction-limited imaging with partially redundant
  masks: II. Optical imaging of faint sources}.
\newline \emph{Optical Society of America Journal}, \textbf{10}(Sept.),
  1882-1894.

\vspace*{0.25cm}
B{\small USCHER}, D.~F., A{\small RMSTRONG}, J.~T., H{\small UMMEL}, C.~A., Q{\small UIRRENBACH}, A.,
  M{\small OZURKEWICH}, D., J{\small OHNSTON}, K.~J., D{\small ENISON}, C.~S., C{\small OLAVITA}, M.~M., \&
  S{\small HAO}, M. 1995.

  \href{http://ukads.nottingham.ac.uk/cgi-bin/nph-bib_query?bibcode=1995ApOpt..34.1081B\&db_key=INST}{Interferometric seeing measurements on Mt. Wilson: power
  spectra and outer scales}.
\newline \emph{Applied Optics}, \textbf{34}(Feb.), 1081-1096.

\vspace*{0.25cm}
C{\small ACCIA}, J.~L., A{\small ZOUIT}, M., \& V{\small ERNIN}, J. 1987.

  \href{http://cdsads.u-strasbg.fr/cgi-bin/nph-bib_query?bibcode=1987ApOpt..26.1288C\&db_key=AST}{Wind and $C_{N}$-squared profiling by single-star
  scintillation analysis}.
\newline \emph{Applied Optics}, \textbf{26}(Apr.), 1288-1294.

\vspace*{0.25cm}
C{\small HELLI}, A. 1987.
 Comparison Between Image Plane Phase Reconstruction Methods in
  Optical Interferometry.
\newline \emph{Revista Mexicana de Astronomia y Astrofisica}, \textbf{14}(May),
  751+.

\vspace*{0.25cm}
C{\small HRISTOU}, J.~C. 1991.

  \href{http://ukads.nottingham.ac.uk/cgi-bin/nph-bib_query?bibcode=1991PASP..103.1040C\&db_key=AST}{Image quality, tip-tilt correction, and shift-and-add
  infrared imaging}.
\newline \emph{Publications of the Astronomical Society of the Pacific}, \textbf{103}(Sept.), 1040-1048.

\vspace*{0.25cm}
C{\small OHEN}, R.~L., G{\small UHATHAKURTA}, P., Y{\small ANNY}, B., S{\small CHNEIDER}, D.~P., \&
  B{\small AHCALL}, J.~N. 1997.

  \href{http://ukads.nottingham.ac.uk/cgi-bin/nph-bib_query?bibcode=1997AJ....113..669C\&db_key=AST}{Globular Cluster Photometry with the Hubble Space
  Telescope.VI.WF/PC-I Observations of the Stellar Populations in the Core of
  M13 (NGC 6205)}.
\newline \emph{Astronomical Journal}, \textbf{113}(Feb.), 669-681.

\vspace*{0.25cm}
C{\small OLAVITA}, M.~M., S{\small HAO}, M., \& S{\small TAELIN}, D.~H. 1987.

  \href{http://ukads.nottingham.ac.uk/cgi-bin/nph-bib_query?bibcode=1987ApOpt..26.4106C\&db_key=INST}{Atmospheric phase measurements with the Mark III stellar
  interferometer}.
\newline \emph{Applied Optics}, \textbf{26}(Oct.), 4106-4112.

\vspace*{0.25cm}
C{\small ONAN}, J., R{\small OUSSET}, G., \& M{\small ADEC}, P. 1995.

  \href{http://ukads.nottingham.ac.uk/cgi-bin/nph-bib_query?bibcode=1995OSAJ...12.1559C\&db_key=INST}{Wave-front temporal spectra in high-resolution imaging
  through turbulence}.
\newline \emph{Optical Society of America Journal}, \textbf{12}(July), 1559-1570.

\vspace*{0.25cm}
C{\small OULMAN}, C.~E., V{\small ERNIN}, J., C{\small OQUEUGNIOT}, Y., \& C{\small ACCIA}, J.-L.
  1988.

  \href{http://ukads.nottingham.ac.uk/cgi-bin/nph-bib_query?bibcode=1988ApOpt..27..155C\&db_key=INST}{Outer scale of turbulence appropriate to modeling
  refractive-index structure profiles}.
\newline \emph{Applied Optics}, \textbf{27}(Jan.), 155+.

\vspace*{0.25cm}
C{\small OX}, A.~N. 2000.
 \emph{Allen's Astrophysical Quantities}.
\newline New York: Springer-Verlag.

\vspace*{0.25cm}
C{\small RAMPTON}, D., M{\small C}C{\small LURE}, R.~D., F{\small LETCHER}, J.~M., \& H{\small UTCHINGS},
  J.~B. 1989.

  \href{http://ukads.nottingham.ac.uk/cgi-bin/nph-bib_query?bibcode=1989AJ.....98.1188C\&db_key=AST}{A search for closely spaced gravitational lenses}.
\newline \emph{Astronomical Journal}, \textbf{98}(Oct.), 1188-1194.

\vspace*{0.25cm}
D{\small AINTY}, J.~C., H{\small ENNINGS}, D.~R., \& O{\small DONNELL}, K.~A. 1981.

  \href{http://ukads.nottingham.ac.uk/cgi-bin/nph-bib_query?bibcode=1981OSAJ...71..490D\&db_key=AST}{Space-time correlation of stellar speckle patterns}.
\newline \emph{Optical Society of America, Journal}, \textbf{71}(Apr.), 490-492.

\vspace*{0.25cm}
D{\small ANTOWITZ}, R.~F. 1998.

  \href{http://ukads.nottingham.ac.uk/cgi-bin/nph-bib_query?bibcode=1998S\%26T....96b..48D\&db_key=AST}{Sharper Images Through Video}.
\newline \emph{Sky and Telescope}, \textbf{96}(Aug.), 48.

\vspace*{0.25cm}
D{\small ANTOWITZ}, R.~F., T{\small EARE}, S.~W., \& K{\small OZUBAL}, M.~J. 2000.

  \href{http://ukads.nottingham.ac.uk/cgi-bin/nph-bib_query?bibcode=2000AJ....119.2455D\&db_key=AST}{Ground-based High-Resolution Imaging of Mercury}.
\newline \emph{Astronomical Journal}, \textbf{119}(May), 2455-2457.

\vspace*{0.25cm}
D{\small AVIS}, J., \& N{\small ORTH}, J.~R. 2001.

  \href{http://ukads.nottingham.ac.uk/cgi-bin/nph-bib_query?bibcode=2001PASA...18..281D\&db_key=AST}{Binary Star Observations in Selected Instants of Good
  Seeing}.
\newline \emph{Publications of the Astronomical Society of Australia}, \textbf{  18}, 281-286.

\vspace*{0.25cm}
D{\small AVIS}, J., L{\small AWSON}, P.~R., B{\small OOTH}, A.~J., T{\small ANGO}, W.~J., \&
  T{\small HORVALDSON}, E.~D. 1995.

  \href{http://ukads.nottingham.ac.uk/cgi-bin/nph-bib_query?bibcode=1995MNRAS.273L..53D\&db_key=AST}{Atmospheric path variations for baselines up to 80m
  measured with the Sydney University Stellar Interferometer}.
\newline \emph{Monthly Notices of the Royal Astronomical Society}, \textbf{273}(Apr.), L53-L58.

\vspace*{0.25cm}
E{\small NGLANDER}, A., S{\small LATKINE}, M., K{\small AROUBI}, R., \& B{\small ENSIMON}, D. 1983.

  \href{http://ukads.nottingham.ac.uk/cgi-bin/nph-bib_query?bibcode=1983OptEn..22..145E\&db_key=INST}{Probabilistic diffraction limited imaging through
  turbulence}.
\newline \emph{Optical Engineering}, \textbf{22}(Feb.), 145-148.

\vspace*{0.25cm}
F{\small RIED}, D.~L. 1965.

  \href{http://ukads.nottingham.ac.uk/cgi-bin/nph-bib_query?bibcode=1965OSAJ...55.1427F\&db_key=AST}{Statistics of a Geometric Representation of Wavefront
  Distortion}.
\newline \emph{Optical Society of America Journal}, \textbf{55}, 1427-1435.

\vspace*{0.25cm}
F{\small RIED}, D.~L. 1978.

  \href{http://ukads.nottingham.ac.uk/cgi-bin/nph-bib_query?bibcode=1978OSAJ...68.1651F\&db_key=INST}{Probability of getting a lucky short-exposure image
  through turbulence}.
\newline \emph{Optical Society of America Journal}, \textbf{68}(Dec.), 1651-1658.

\vspace*{0.25cm}
F{\small RISCH}, U., S{\small ULEM}, P., \& N{\small ELKIN}, M. 1978.
 A simple dynamical model of intermittent fully developed
  turbulence.
\newline \emph{Journal of Fluid Mechancis}, \textbf{87}, 719-736.

\vspace*{0.25cm}
G{\small RAVES}, J.~E., N{\small ORTHCOTT}, M.~J., R{\small ODDIER}, F.~J., R{\small ODDIER}, C.~A.,
  \& C{\small LOSE}, L.~M. 1998 (Sept.).

  \href{http://ukads.nottingham.ac.uk/cgi-bin/nph-bib_query?bibcode=1998SPIE.3353...34G\&db_key=AST}{First light for Hokupa'a: 36-element curvature AO system
  at UH}.
\newline \emph{Pages  34-43 of:} \emph{SPIE Proceedings, Vol. 3353, Adaptive Optical
  System Technologies, D. Bonaccini, R. Tyson; Eds.}

\vspace*{0.25cm}
G{\small UHATHAKURTA}, P., Y{\small ANNY}, B., S{\small CHNEIDER}, D.~P., \& B{\small AHCALL}, J.~N.
  1996.

  \href{http://cdsads.u-strasbg.fr/cgi-bin/nph-bib_query?bibcode=1996AJ....111..267G\&db_key=AST}{Globular Cluster Photometry With the Hubble Space Telescope.
  V. WFPC Study of M15's Central density Cusp}.
\newline \emph{Astronomical Journal}, \textbf{111}(Jan.), 267+.

\vspace*{0.25cm}
H{\small ANIFF}, C.~A., \& B{\small USCHER}, D.~F. 1992.

  \href{http://ukads.nottingham.ac.uk/cgi-bin/nph-bib_query?bibcode=1992OSAJ....9..203H\&db_key=INST}{Diffraction-limited imaging with partially redundant
  masks. I. Infrared imaging of bright objects}.
\newline \emph{Optical Society of America Journal}, \textbf{9}(Feb.), 203-218.

\vspace*{0.25cm}
H{\small ANIFF}, C.~A., M{\small ACKAY}, C.~D., T{\small ITTERINGTON}, D.~J., S{\small IVIA}, D., \&
  B{\small ALDWIN}, J.~E. 1987.

  \href{http://ukads.nottingham.ac.uk/cgi-bin/nph-bib_query?bibcode=1987Natur.328..694H\&db_key=AST}{The first images from optical aperture synthesis}.
\newline \emph{Nature}, \textbf{328}(Aug.), 694-696.

\vspace*{0.25cm}
H{\small ANIFF}, C.~A., B{\small USCHER}, D.~F., C{\small HRISTOU}, J.~C., \& R{\small IDGWAY}, S.~T.
  1989.

  \href{http://ukads.nottingham.ac.uk/cgi-bin/nph-bib_query?bibcode=1989MNRAS.241P..51H\&db_key=AST}{Synthetic aperture imaging at infrared wavelengths}.
\newline \emph{Monthly Notices of the Royal Astronomical Society}, \textbf{241}(Nov.), 51P-56PP.

\vspace*{0.25cm}
H{\small ARDY}, J.~W. 1998.
 \emph{Adaptive optics for astronomical telescopes}.
\newline New York: Oxford University Press.

\vspace*{0.25cm}
H{\small ARRIS}, E.~J., R{\small OYLE}, G.~J., S{\small PENCER}, R.~D., S{\small PENCER}, S., \&
  S{\small USKE}, W. 2000.
 Evaluation of a Novel CCD Camera for Dose Reduction in Digital
  Radiography.
\newline \emph{Proceedings of the IEEE Nuclear Science Symposium and Medical
  Imaging Conference, Lyons}, Sept.

\vspace*{0.25cm}
H{\small ECQUET}, J., \& C{\small OUPINOT}, G. 1985.

  \href{http://ukads.nottingham.ac.uk/cgi-bin/nph-bib_query?bibcode=1985JOpt...16...21H\&db_key=INST}{A gain in resolution by the superposition of selected
  recentered short exposures}.
\newline \emph{Journal of Optics}, \textbf{16}(Feb.), 21-26.

\vspace*{0.25cm}
H{\small YNECEK}, J., \& N{\small ISHIWAKI}, T. 2002 (July).
 Recent progress toward single photon detection using charge
  multiplying CCD image sensors.
\newline \emph{Pages  7-12 of:} \emph{Proc. 16th World Multiconference on
  Systems and Cybernetics, SCI2002/ISAS2002, No. 2}.

\vspace*{0.25cm}
I{\small SHIMARU}, A. 1978.
 \emph{Wave propagation and scattering in random media, Vol. 2:
  Multiple scattering, turbulence, rough surfaces and remote sensing}.
\newline New York ; London: Academic Press.

\vspace*{0.25cm}
J{\small ERRAM}, P., P{\small OOL}, P.~J., B{\small ELL}, R., B{\small URT}, D.~J., B{\small OWRING}, S.,
  S{\small PENCER}, S., H{\small AZELWOOD}, M., M{\small OODY}, I., C{\small ATLETT}, N., \& H{\small EYES},
  P.~S. 2001 (May).

  \href{http://ukads.nottingham.ac.uk/cgi-bin/nph-bib_query?bibcode=2001SPIE.4306..178J\&db_key=INST}{The LLCCD: low-light imaging without the need for an
  intensifier}.
\newline \emph{Pages  178-186 of:} \emph{SPIE Proceedings, Vol. 4306, Sensors and
  Camera Systems II, M. Blouke et al Eds.}

\vspace*{0.25cm}
K{\small ARO}, D.~P., \& S{\small CHNEIDERMAN}, A.~M. 1978.

  \href{http://ukads.nottingham.ac.uk/cgi-bin/nph-bib_query?bibcode=1978OSAJ...68..480K\&db_key=INST}{Speckle interferometry at finite spectral bandwidths and
  exposure times (E)}.
\newline \emph{Optical Society of America Journal}, \textbf{68}, 480-+.

\vspace*{0.25cm}
K{\small ARR}, T.~J. 1991.

  \href{http://ukads.nottingham.ac.uk/cgi-bin/nph-bib_query?bibcode=1991ApOpt..30..363K\&db_key=INST}{Temporal response of atmospheric turbulence
  compensation}.
\newline \emph{Applied Optics}, \textbf{30}(Feb.), 363+.

\vspace*{0.25cm}
K{\small EEN}, J.~W. 1999 (Dec.).
 \emph{Personal communication - algorithm for generating Kolmogorov
  turbulence developed by Dave Buscher and modified by James Keen}.

\vspace*{0.25cm}
K{\small ENWORTHY}, M., H{\small OFMANN}, K., C{\small LOSE}, L., H{\small INZ}, P., M{\small AMAJEK}, E.,
  S{\small CHERTL}, D., W{\small EIGELT}, G., A{\small NGEL}, R., B{\small ALEGA}, Y.~Y., H{\small INZ}, J., \&
  R{\small IEKE}, G. 2001.

  \href{http://cdsads.u-strasbg.fr/cgi-bin/nph-bib_query?bibcode=2001ApJ...554L..67K\&db_key=AST}{Gliese 569B: A Young Multiple Brown Dwarf System?}
\newline \emph{Astrophysical Journal Letters}, \textbf{554}(June),
  L67-LL70 (\href{http://www.arxiv.org/abs/astro-ph/0105157}{astro-ph/0105157}).

\vspace*{0.25cm}
K{\small IM}, Y., \& J{\small AGGARD}, D.~L. 1988.

  \href{http://ukads.nottingham.ac.uk/cgi-bin/nph-bib_query?bibcode=1988OSAJ....5..475K\&db_key=INST}{Band-limited fractal model of atmospheric refractivity
  fluctuation}.
\newline \emph{Optical Society of America Journal}, \textbf{5}(Apr.), 475-480.

\vspace*{0.25cm}
K{\small OLMOGOROV}, A.~N. 1941a.
 Dissipation of energy in the locally isotropic turbulence.
\newline \emph{Comptes rendus (Doklady) de l'Académie des Sciences de
  l'U.R.S.S.}, \textbf{32}, 16-18.

\vspace*{0.25cm}
K{\small OLMOGOROV}, A.~N. 1941b.
 The local structure of turbulence in incompressible viscous fluid
  for very large Reynold's numbers.
\newline \emph{Comptes rendus (Doklady) de l'Académie des Sciences de
  l'U.R.S.S.}, \textbf{30}, 301-305.

\vspace*{0.25cm}
K{\small UO}, A.~Y.-S., \& C{\small ORRSIN}, S. 1972.
 \emph{Journal of Fluid Mechancis}, \textbf{56}, 447.

\vspace*{0.25cm}
L{\small ABEYRIE}, A. 1970.

  \href{http://cdsads.u-strasbg.fr/cgi-bin/nph-bib_query?bibcode=1970A\%26A.....6...85L\&db_key=AST}{Attainment of Diffraction Limited Resolution in Large
  Telescopes by Fourier Analysing Speckle Patterns in Star Images}.
\newline \emph{Astronomy and Astrophysics}, \textbf{6}(May), 85+.

\vspace*{0.25cm}
L{\small ELIEVRE}, G., N{\small IETO}, J.-L., T{\small HOUVENOT}, E., S{\small ALMON}, D., \&
  L{\small LEBARIA}, A. 1988.

  \href{http://ukads.nottingham.ac.uk/cgi-bin/nph-bib_query?bibcode=1988A\%26A...200..301L\&db_key=AST}{Very high resolution imaging using sub-pupil
  apertures, recentering and selection of short exposures}.
\newline \emph{Astronomy and Astrophysics}, \textbf{200}(July), 301-311.

\vspace*{0.25cm}
L{\small INFIELD}, R.~P., C{\small OLAVITA}, M.~M., \& L{\small ANE}, B.~F. 2001.

  \href{http://ukads.nottingham.ac.uk/cgi-bin/nph-bib_query?bibcode=2001ApJ...554..505L\&db_key=AST}{Atmospheric Turbulence Measurements with the Palomar
  Testbed Interferometer}.
\newline \emph{Astrophysical Journal}, \textbf{554}(June), 505-513 (\href{http://www.arxiv.org/abs/astro-ph/0102052}{astro-ph/0102052}).

\vspace*{0.25cm}
L{\small OHMANN}, A.~W., \& W{\small EIGELT}, G.~P. 1979.
 Astronomical speckle interferometry; measurements of isoplanicity
  and of temporal correlation.
\newline \emph{Optik}, \textbf{53}(Aug.), 167-180.

\vspace*{0.25cm}
L{\small OPEZ}, B., \& S{\small ARAZIN}, M. 1993.

  \href{http://ukads.nottingham.ac.uk/cgi-bin/nph-bib_query?bibcode=1993A\%26A...276..320L\&db_key=AST}{The ESO atmospheric temporal coherence monitor
  dedicated to high angular resolution imaging}.
\newline \emph{Astronomy and Astrophysics}, \textbf{276}(Sept.), 320+.

\vspace*{0.25cm}
M{\small ACKAY}, C.~D., T{\small UBBS}, R.~N., B{\small ELL}, R., B{\small URT}, D.~J., J{\small ERRAM}, P.,
  \& M{\small OODY}, I. 2001 (May).

  \href{http://ukads.nottingham.ac.uk/cgi-bin/nph-bib_query?bibcode=2001SPIE.4306..289M\&db_key=INST}{Subelectron read noise at MHz pixel rates}.
\newline \emph{Pages  289-298 of:} \emph{SPIE Proceedings, Vol. 4306, Sensors and
  Camera Systems II, M. M. Blouke et al Eds.} (\href{http://www.arxiv.org/abs/astro-ph/0101409}{astro-ph/0101409}).

\vspace*{0.25cm}
M{\small ANDELBROT}, B.~B. 1974.
 Intermittent turbulence in self-similar cascades: divergence of high
  moments and dimension of the carrier.
\newline \emph{Journal of Fluid Mechancis}, \textbf{62}, 331-358.

\vspace*{0.25cm}
M{\small ARKS}, R.~D., V{\small ERNIN}, J., A{\small ZOUIT}, M., M{\small ANIGAULT}, J.~F., \&
  C{\small LEVELIN}, C. 1999.

  \href{http://ukads.nottingham.ac.uk/cgi-bin/nph-bib_query?bibcode=1999A..134..161M\&db_key=AST}{Measurement of optical seeing on the high antarctic
  plateau}.
\newline \emph{Astronomy and Astrophysics Supplement Series}, \textbf{134}(Jan.), 161-172.

\vspace*{0.25cm}
M{\small ARTIN}, F., C{\small ONAN}, R., T{\small OKOVININ}, A., Z{\small IAD}, A., T{\small RINQUET}, H.,
  B{\small ORGNINO}, J., A{\small GABI}, A., \& S{\small ARAZIN}, M. 2000.

  \href{http://ukads.nottingham.ac.uk/cgi-bin/nph-bib_query?bibcode=2000A\%26AS..144...39M\&db_key=AST}{Optical parameters relevant for High Angular
  Resolution at Paranal from GSM instrument and surface layer contribution}.
\newline \emph{Astronomy and Astrophysics Supplement Series}, \textbf{144}(May), 39-44.

\vspace*{0.25cm}
M{\small C}A{\small LISTER}, H.~A., H{\small ARTKOPF}, W.~I., S{\small OWELL}, J.~R., D{\small OMBROWSKI},
  E.~G., \& F{\small RANZ}, O.~G. 1989.

  \href{http://cdsads.u-strasbg.fr/cgi-bin/nph-bib_query?bibcode=1989AJ.....97..510M\&db_key=AST}{ICCD speckle observations of binary stars. IV -
  Measurements during 1986-1988 from the Kitt Peak 4 M telescope}.
\newline \emph{Astronomical Journal}, \textbf{97}(Feb.), 510-531.

\vspace*{0.25cm}
M{\small ONNIER}, J.~D. 2003.

  \href{http://ukads.nottingham.ac.uk/cgi-bin/nph-bib_query?bibcode=2003RPPh...66..789M\&db_key=AST}{Optical interferometry in astronomy}.
\newline \emph{Reports of Progress in Physics}, \textbf{66}(May), 789-857
  (\href{http://www.arxiv.org/abs/astro-ph/0307036}{astro-ph/0307036}).

\vspace*{0.25cm}
M{\small U{\~ N}OZ-}T{\small U{\~ N}{\' O}N}, C., V{\small ERNIN}, J., \& V{\small ARELA}, A.~M.
  1997.

  \href{http://ukads.nottingham.ac.uk/cgi-bin/nph-bib_query?bibcode=1997A\%26AS..125..183M\&db_key=AST}{Night-time image quality at Roque de los Muchachos
  Observatory}.
\newline \emph{Astronomy and Astrophysics Supplement Series}, \textbf{125}(Oct.), 183-193.

\vspace*{0.25cm}
N{\small AHMAN}, N.~S., \& G{\small UILLAUME}, M.~E. 1981.
 Deconvolution of time domain waveforms in the presence of noise.
\newline \emph{NBS Technical Note}, \textbf{1047}(Oct.), 1-122.

\vspace*{0.25cm}
N{\small IETO}, J.-L., \& T{\small HOUVENOT}, E. 1991.

  \href{http://ukads.nottingham.ac.uk/cgi-bin/nph-bib_query?bibcode=1991A\%26A...241..663N\&db_key=AST}{Recentring and selection of short-exposure images with
  photon-counting detectors. I - Reliability tests}.
\newline \emph{Astronomy and Astrophysics}, \textbf{241}(Jan.), 663-672.

\vspace*{0.25cm}
N{\small IETO}, J.-L., L{\small LEBARIA}, A., \& {\small DI }S{\small EREGO }A{\small LIGHIERI}, S. 1987.

  \href{http://adsabs.harvard.edu/cgi-bin/nph-bib_query?bibcode=1987A\%26A...178..301N\&db_key=AST}{Photon-counting detectors in time-resolved imaging mode -
  Image recentring and selection algorithms}.
\newline \emph{Astronomy and Astrophysics}, \textbf{178}(May), 301-306.

\vspace*{0.25cm}
N{\small IETO}, J.-L., R{\small OQUES}, S., L{\small LEBARIA}, A., V{\small ANDERRIEST}, C.,
  L{\small ELIEVRE}, G., {\small DI }S{\small EREGO }A{\small LIGHIERI}, S., M{\small ACCHETTO}, F.~D., \& P{\small ERRYMAN},
  M.~A.~C. 1988.

  \href{http://adsabs.harvard.edu/cgi-bin/nph-bib_query?bibcode=1988ApJ...325..644N\&db_key=AST}{High-resolution imaging of the double QSO 2345 + 007 -
  Evidence for subcomponents}.
\newline \emph{Astrophysical Journal}, \textbf{325}(Feb.), 644-650.

\vspace*{0.25cm}
N{\small IETO}, J.-L., A{\small URIERE}, M., S{\small EBAG}, J., A{\small RNAUD}, J., L{\small ELIEVRE}, G.,
  B{\small LAZIT}, A., F{\small OY}, R., B{\small ONALDO}, S., \& T{\small HOUVENOT}, E. 1990.

  \href{http://ukads.nottingham.ac.uk/cgi-bin/nph-bib_query?bibcode=1990A\%26A...239..155N\&db_key=AST}{The optical counterpart of the X-ray binary in the
  globular cluster NGC 6712}.
\newline \emph{Astronomy and Astrophysics}, \textbf{239}(Nov.), 155-162.

\vspace*{0.25cm}
N{\small IGHTINGALE}, N.~S., \& B{\small USCHER}, D.~F. 1991.

  \href{http://ukads.nottingham.ac.uk/cgi-bin/nph-bib_query?bibcode=1991MNRAS.251..155N\&db_key=AST}{Interferometric seeing measurements at the La Palma
  Observatory}.
\newline \emph{Monthly Notices of the Royal Astronomical Society}, \textbf{251}(July), 155-166.

\vspace*{0.25cm}
N{\small OLL}, R.~J. 1976.

  \href{http://ukads.nottingham.ac.uk/cgi-bin/nph-bib_query?bibcode=1976OSAJ...66..207N\&db_key=INST}{Zernike polynomials and atmospheric turbulence}.
\newline \emph{Optical Society of America Journal}, \textbf{66}(Mar.), 207-211.

\vspace*{0.25cm}
O'B{\small YRNE}, J.~W. 1988.

  \href{http://ukads.nottingham.ac.uk/cgi-bin/nph-bib_query?bibcode=1988PASP..100.1169O\&db_key=AST}{Seeing measurements using a shearing interferometer}.
\newline \emph{Publications of the Astronomical Society of the Pacific}, \textbf{100}(Sept.), 1169-1177.

\vspace*{0.25cm}
P{\small ARRY}, G., W{\small ALKER}, J.~G., \& S{\small CADDAN}, R.~J. 1979.
 Statistics of stellar speckle patterns and pupil plane
  scintillation.
\newline \emph{Optica Acta}, \textbf{26}(Oct.), 563.

\vspace*{0.25cm}
P{\small OGSON}, N. 1856.

  \href{http://ukads.nottingham.ac.uk/cgi-bin/nph-bib_query?bibcode=1856MNRAS..17...12P\&db_key=AST}{Magnitudes of Thirty-six of the Minor Planets for the
  first day of each month of the year 1857}.
\newline \emph{Monthly Notices of the Royal Astronomical Society}, \textbf{17}(Nov.), 12-15.

\vspace*{0.25cm}
R{\small ACINE}, R. 1996.

  \href{http://ukads.nottingham.ac.uk/cgi-bin/nph-bib_query?bibcode=1996PASP..108..372R\&db_key=AST}{Temporal Fluctuations of Atmospheric Seeing}.
\newline \emph{Publications of the Astronomical Society of the Pacific}, \textbf{108}(Apr.), 372+.

\vspace*{0.25cm}
R{\small AGAZZONI}, R., \& F{\small ARINATO}, J. 1999.

  \href{http://ukads.nottingham.ac.uk/cgi-bin/nph-bib_query?bibcode=1999A\%26A...350L..23R\&db_key=AST}{Sensitivity of a pyramidic Wave Front sensor in closed
  loop Adaptive Optics}.
\newline \emph{Astronomy and Astrophysics}, \textbf{350}(Oct.), L23-L26.

\vspace*{0.25cm}
R{\small OBBINS}, M.~S., \& H{\small ADWEN}, B.~J. 2003.
 The noise performance of electron multiplying charge coupled
  devices.
\newline \emph{IEEE Transactions on Electron Devices}, \textbf{50}(May),
  1227-1232.

\vspace*{0.25cm}
R{\small ODDIER}, F. 1981.

  \href{http://ukads.nottingham.ac.uk/cgi-bin/nph-bib_query?bibcode=1981prop...19..281R\&db_key=AST}{The effects of atmospheric turbulence in optical
  astronomy}.
\newline \emph{Pages  281-376 of:} \emph{Progress in optics. Volume 19.
  Amsterdam, North-Holland Publishing Co.}

\vspace*{0.25cm}
R{\small ODDIER}, F. 1988.

  \href{http://ukads.nottingham.ac.uk/cgi-bin/nph-bib_query?bibcode=1988PhR...170...99R\&db_key=AST}{Interferometric imaging in optical astronomy}.
\newline \emph{Physics Reports}, \textbf{170}, 99-166.

\vspace*{0.25cm}
R{\small ODDIER}, F., G{\small ILLI}, J.~M., \& L{\small UND}, G. 1982a.

  \href{http://ukads.nottingham.ac.uk/cgi-bin/nph-bib_query?bibcode=1982JOpt...13..263R\&db_key=AST}{On the origin of speckle boiling and its effects in
  stellar speckle interferometry}.
\newline \emph{Journal of Optics}, \textbf{13}(Oct.), 263-271.

\vspace*{0.25cm}
R{\small ODDIER}, F., G{\small ILLI}, J.~M., \& V{\small ERNIN}, J. 1982b.

  \href{http://ukads.nottingham.ac.uk/cgi-bin/nph-bib_query?bibcode=1982JOpt...13...63R\&db_key=INST}{On the isoplanatic patch size in stellar speckle
  interferometry}.
\newline \emph{Journal of Optics}, \textbf{13}(Mar.), 63-70.

\vspace*{0.25cm}
R{\small ODDIER}, F., C{\small OWIE}, L., G{\small RAVES}, J.~E., S{\small ONGAILA}, A., \& M{\small C}K{\small ENNA},
  D. 1990 (July).

  \href{http://ukads.nottingham.ac.uk/cgi-bin/nph-bib_query?bibcode=1990SPIE.1236..485R\&db_key=AST}{Seeing at Mauna Kea - A joint UH-UN-NOAO-CFHT study}.
\newline \emph{Pages  485-491 of:} \emph{SPIE Proceedings, Vol. 1236, Advanced
  technology at optical telescopes IV Part 1}.

\vspace*{0.25cm}
R{\small ODDIER}, F., N{\small ORTHCOTT}, M.~J., G{\small RAVES}, J.~E., M{\small C}K{\small ENNA}, D.~L., \&
  R{\small ODDIER}, D. 1993.

  \href{http://ukads.nottingham.ac.uk/cgi-bin/nph-bib_query?bibcode=1993OSAJ...10..957R\&db_key=INST}{One-dimensional spectra of turbulence-induced Zernike
  aberrations: time-delay and isoplanicity error in partial adaptive
  compensation}.
\newline \emph{Optical Society of America Journal}, \textbf{10}(May), 957-965.

\vspace*{0.25cm}
S{\small AINT-}J{\small ACQUES}, D., \& B{\small ALDWIN}, J.~E. 2000 (July).

  \href{http://ukads.nottingham.ac.uk/cgi-bin/nph-bib_query?bibcode=2000SPIE.4006..951S\&db_key=AST}{Taylor's hypothesis: good for nuts}.
\newline \emph{Pages  951-962 of:} \emph{SPIE Proceedings, Vol. 4006, Interferometry
  in Optical Astronomy, P. Lena; A. Quirrenbach; Eds.}

\vspace*{0.25cm}
S{\small AINT-}J{\small ACQUES}, D., C{\small OX}, G.~C., B{\small ALDWIN}, J.~E., M{\small ACKAY}, C.~D.,
  W{\small ALDRAM}, E.~M., \& W{\small ILSON}, R.~W. 1997.

  \href{http://ukads.nottingham.ac.uk/cgi-bin/nph-bib_query?bibcode=1997MNRAS.290...66S\&db_key=AST}{The JOSE atmospheric seeing monitor at the William
  Herschel Telescope}.
\newline \emph{Monthly Notices of the Royal Astronomical Society}, \textbf{290}(Sept.), 66-74.

\vspace*{0.25cm}
S{\small CADDAN}, R.~J., \& W{\small ALKER}, J.~G. 1978.

  \href{http://ukads.nottingham.ac.uk/cgi-bin/nph-bib_query?bibcode=1978ApOpt..17.3779S\&db_key=AST}{Statistics of stellar speckle patterns}.
\newline \emph{Applied Optics}, \textbf{17}(Dec.), 3779-3784.

\vspace*{0.25cm}
S{\small IGGIA}, E.~D. 1978.

  \href{http://ukads.nottingham.ac.uk/cgi-bin/nph-bib_query?bibcode=1978PhRvA..17.1166S\&db_key=INST}{Model of intermittency in three-dimensional
  turbulence}.
\newline \emph{Physical Review A}, \textbf{17}(Mar.), 1166-1176.

\vspace*{0.25cm}
S{\small {\O}RENSEN}, A.~N. 2002 (Feb.).
 \emph{Personal communication - the measured shape of aberrations in
  the NOT optics}.

\vspace*{0.25cm}
S{\small TREHL}, K. 1895.
 Aplanatische und fehlerhafte Abbildung im Fernrohr.
\newline \emph{Zeitschrift für Instrumentenkunde}, \textbf{15}(Oct.),
  362-370.

\vspace*{0.25cm}
S{\small TREHL}, K. 1902.
 Ueber Luftschlieren und Zonenfehler.
\newline \emph{Zeitschrift für Instrumentenkunde}, \textbf{22}(July), 213-217.

\vspace*{0.25cm}
T{\small ATARSKI}, V.~I. 1961.
 \emph{Wave Propagation in a Turbulent Medium}.
\newline McGraw-Hill.

\vspace*{0.25cm}
T{\small AYLOR}, G.~I. 1938 (Feb.).
 The spectrum of turbulence.
\newline \emph{Pages  476-490 of:} \emph{Proceedings of the Royal Society of
  London A, 164}.

\vspace*{0.25cm}
T{\small UBBS}, R.~N., B{\small ALDWIN}, J.~E., M{\small ACKAY}, C.~D., \& C{\small OX}, G.~C. 2002.

  \href{http://ukads.nottingham.ac.uk/cgi-bin/nph-bib_query?bibcode=2002A\%26A...387L..21T\&db_key=AST}{Diffraction-limited CCD imaging with faint reference
  stars}.
\newline \emph{Astronomy and Astrophysics}, \textbf{387}(May), L21-L24 (\href{http://www.arxiv.org/abs/astro-ph/0203470}{astro-ph/0203470}).

\vspace*{0.25cm}
T{\small UTHILL}, P.~G., M{\small ONNIER}, J.~D., D{\small ANCHI}, W.~C., W{\small ISHNOW}, E.~H., \&
  H{\small ANIFF}, C.~A. 2000.

  \href{http://ukads.nottingham.ac.uk/cgi-bin/nph-bib_query?bibcode=2000PASP..112..555T\&db_key=AST}{Michelson Interferometry with the Keck I Telescope}.
\newline \emph{Publications of the Astronomical Society of the Pacific},
  \textbf{112}(Apr.), 555-565 (\href{http://www.arxiv.org/abs/astro-ph/0003146}{astro-ph/0003146}).

\vspace*{0.25cm}
V{\small ERNIN}, J., \& M{\small U{\~ N}OZ-}T{\small U{\~ N}{\' O}N}, C. 1994.

  \href{http://ukads.nottingham.ac.uk/cgi-bin/nph-bib_query?bibcode=1994A\%26A...284..311V\&db_key=AST}{Optical seeing at La Palma Observatory. 2: Intensive
  site testing campaign at the Nordic Optical Telescope}.
\newline \emph{Astronomy and Astrophysics}, \textbf{284}(Apr.), 311-318.

\vspace*{0.25cm}
V{\small ERNIN}, J., \& M{\small U{\~ N}OZ-}T{\small U{\~ N}{\' O}N}, C. 1998.

  \href{http://cdsads.u-strasbg.fr/cgi-bin/nph-bib_query?bibcode=1998NewAR..42..451V\&db_key=AST}{The temporal behaviour of seeing}.
\newline \emph{New Astronomy Review}, \textbf{42}(Nov.), 451-454.

\vspace*{0.25cm}
V{\small ERNIN}, J., \& R{\small ODDIER}, F. 1973.
 Experimental determination of two-dimensional spatiotemporal power
  spectra of stellar light scintillation. Evidence for a multilayer structure
  of the air turbulence in the upper troposphere.
\newline \emph{Journal of the Optical Society of America}, \textbf{63}(Mar.),
  270-273.

\vspace*{0.25cm}
V{\small ERNIN}, J., C{\small ACCIA}, J.-L., W{\small EIGELT}, G., \& M{\small UELLER}, M. 1991.

  \href{http://ukads.nottingham.ac.uk/cgi-bin/nph-bib_query?bibcode=1991A\%26A...243..553V\&db_key=AST}{Speckle lifetime and isoplanicity determinations -
  Direct measurements and derivation from turbulence and wind profiles}.
\newline \emph{Astronomy and Astrophysics}, \textbf{243}(Mar.), 553-558.

\vspace*{0.25cm}
W{\small ILSON}, R.~W. 2002.

  \href{http://ukads.nottingham.ac.uk/cgi-bin/nph-bib_query?bibcode=2002MNRAS.337..103W\&db_key=AST}{SLODAR: measuring optical turbulence altitude with a
  Shack-Hartmann wavefront sensor}.
\newline \emph{Monthly Notices of the Royal Astronomical Society}, \textbf{337}(Nov.), 103-108.

\vspace*{0.25cm}
W{\small ILSON}, R.~W. 2003 (July).
 \emph{Personal communication - results from the ING seeing monitor
  at Roque de los Muchachos Observatory}.

\vspace*{0.25cm}
W{\small ILSON}, R.~W., \& S{\small AUNTER}, C. 2003 (Feb.).

  \href{http://ukads.nottingham.ac.uk/cgi-bin/nph-bib_query?bibcode=2003SPIE.4839..466W\&db_key=INST}{Turbulence profiler and seeing monitor for laser guide
  star adaptive optics}.
\newline \emph{Pages  466-472 of:} \emph{SPIE Proceedings, Vol. 4839, Adaptive Optical
  System Technologies II, P. Wizinowich and D. Bonaccini Eds.}

\vspace*{0.25cm}
W{\small ILSON}, R.~W., O'M{\small AHONY}, N., P{\small ACKHAM}, C., \& A{\small ZZARO}, M. 1999.

  \href{http://ukads.nottingham.ac.uk/cgi-bin/nph-bib_query?bibcode=1999MNRAS.309..379W\&db_key=AST}{The seeing at the William Herschel Telescope}.
\newline \emph{Monthly Notices of the Royal Astronomical Society}, \textbf{309}(Oct.), 379-387.

\vspace*{0.25cm}
Y{\small OUNG}, J.~S., B{\small ALDWIN}, J.~E., B{\small OYSEN}, R.~C., H{\small ANIFF}, C.~A.,
  L{\small AWSON}, P.~R., M{\small ACKAY}, C.~D., P{\small EARSON}, D., R{\small OGERS}, J., S{\small T.-}J{\small ACQUES},
  D., W{\small ARNER}, P.~J., W{\small ILSON}, D.~M.~A., \& W{\small ILSON}, R.~W. 2000.

  \href{http://ukads.nottingham.ac.uk/cgi-bin/nph-bib_query?bibcode=2000MNRAS.315..635Y\&db_key=AST}{New views of Betelgeuse: multi-wavelength surface imaging
  and implications for models of hotspot generation}.
\newline \emph{Monthly Notices of the Royal Astronomical Society}, \textbf{315}(July), 635-645.

\vspace*{0.25cm}
Y{\small OUNG}, J.~S., B{\small ALDWIN}, J.~E., B{\small ASDEN}, A.~G., B{\small HARMAL}, N.~A.,
  B{\small USCHER}, D.~F., G{\small EORGE}, A.~V., H{\small ANIFF}, C.~A., K{\small EEN}, J.~W.,
  O'D{\small ONOVAN}, B., P{\small EARSON}, D., T{\small HORSTEINSSON}, H., T{\small HUREAU}, N.~D.,
  T{\small UBBS}, R.~N., \& W{\small ARNER}, P.~J. 2003 (Feb.).

  \href{http://ukads.nottingham.ac.uk/cgi-bin/nph-bib_query?bibcode=2003SPIE.4838..369Y\&db_key=INST}{Astrophysical results from COAST}.
\newline \emph{Pages  369-378 of:} \emph{SPIE Proceedings, Vol. 4838, Interferometry
  for Optical Astronomy II, W. Traub Eds.}

\vspace*{0.25cm}
Z{\small IAD}, A., B{\small ORNINO}, J., M{\small ARTIN}, F., \& A{\small GABI}, A. 1994.

  \href{http://ukads.nottingham.ac.uk/cgi-bin/nph-bib_query?bibcode=1994A\%26A...282.1021Z\&db_key=AST}{Experimental estimation of the spatial-coherence outer
  scale from a wavefront statistical analysis}.
\newline \emph{Astronomy and Astrophysics}, \textbf{282}(Feb.), 1021-1033.

\end{document}